\documentclass[twocolumn,amsmath,showpacs,prl,superscriptaddress]{revtex4}

\usepackage{graphicx}
\usepackage{dcolumn} 
\usepackage{bm} 
\usepackage[dvips]{color}

\def\ARNPS{ Ann. Rev. Nucl. Part. Sci.}
\def\PPNP{ Prog. Part. Nucl. Phys.}

\def\PLB{{ Phys. Lett.}  B}
\def\NPA{{ Nucl. Phys.}  A}
\def\NPB{{ Nucl. Phys.}  B}
\def\PRL{ Phys. Rev. Lett.}
\def\PRC{{ Phys. Rev.} C}

\def\pT{\mbox{$p_T$}}
\def\pTtrig{\mbox{$p^{\rm trig}_T$}}
\def\pTassoc{\mbox{$p^{\rm assoc}_T$}}
\def\deta{\mbox{$\Delta\eta$}}
\def\dphi{\mbox{$\Delta\phi$}}

\def\Journal#1#2#3#4{{#1} {\bf #2}, #3 (#4)}

\begin{document}

\title{Direct observation of dijets in central Au+Au collisions at $\sqrt{s_{NN}}=200$ GeV}

\date{\today}

\affiliation{Argonne National Laboratory, Argonne, Illinois 60439}
\affiliation{University of Birmingham, Birmingham, United Kingdom}
\affiliation{Brookhaven National Laboratory, Upton, New York 11973}
\affiliation{California Institute of Technology, Pasadena, California 91125}
\affiliation{University of California, Berkeley, California 94720}
\affiliation{University of California, Davis, California 95616}
\affiliation{University of California, Los Angeles, California 90095}
\affiliation{Carnegie Mellon University, Pittsburgh, Pennsylvania 15213}
\affiliation{Creighton University, Omaha, Nebraska 68178}
\affiliation{Nuclear Physics Institute AS CR, 250 68 \v{R}e\v{z}/Prague, Czech Republic}
\affiliation{Laboratory for High Energy (JINR), Dubna, Russia}
\affiliation{Particle Physics Laboratory (JINR), Dubna, Russia}
\affiliation{University of Frankfurt, Frankfurt, Germany}
\affiliation{Institute of Physics, Bhubaneswar 751005, India}
\affiliation{Indian Institute of Technology, Mumbai, India}
\affiliation{Indiana University, Bloomington, Indiana 47408}
\affiliation{Institut de Recherches Subatomiques, Strasbourg, France}
\affiliation{University of Jammu, Jammu 180001, India}
\affiliation{Kent State University, Kent, Ohio 44242}
\affiliation{Institute of Modern Physics, Lanzhou, China}
\affiliation{Lawrence Berkeley National Laboratory, Berkeley, California 94720}
\affiliation{Massachusetts Institute of Technology, Cambridge, MA 02139-4307}
\affiliation{Max-Planck-Institut f\"ur Physik, Munich, Germany}
\affiliation{Michigan State University, East Lansing, Michigan 48824}
\affiliation{Moscow Engineering Physics Institute, Moscow Russia}
\affiliation{City College of New York, New York City, New York 10031}
\affiliation{NIKHEF and Utrecht University, Amsterdam, The Netherlands}
\affiliation{Ohio State University, Columbus, Ohio 43210}
\affiliation{Panjab University, Chandigarh 160014, India}
\affiliation{Pennsylvania State University, University Park, Pennsylvania 16802}
\affiliation{Institute of High Energy Physics, Protvino, Russia}
\affiliation{Purdue University, West Lafayette, Indiana 47907}
\affiliation{Pusan National University, Pusan, Republic of Korea}
\affiliation{University of Rajasthan, Jaipur 302004, India}
\affiliation{Rice University, Houston, Texas 77251}
\affiliation{Universidade de Sao Paulo, Sao Paulo, Brazil}
\affiliation{University of Science \& Technology of China, Hefei 230026, China}
\affiliation{Shanghai Institute of Applied Physics, Shanghai 201800, China}
\affiliation{SUBATECH, Nantes, France}
\affiliation{Texas A\&M University, College Station, Texas 77843}
\affiliation{University of Texas, Austin, Texas 78712}
\affiliation{Tsinghua University, Beijing 100084, China}
\affiliation{Valparaiso University, Valparaiso, Indiana 46383}
\affiliation{Variable Energy Cyclotron Centre, Kolkata 700064, India}
\affiliation{Warsaw University of Technology, Warsaw, Poland}
\affiliation{University of Washington, Seattle, Washington 98195}
\affiliation{Wayne State University, Detroit, Michigan 48201}
\affiliation{Institute of Particle Physics, CCNU (HZNU), Wuhan 430079, China}
\affiliation{Yale University, New Haven, Connecticut 06520}
\affiliation{University of Zagreb, Zagreb, HR-10002, Croatia}

\author{J.~Adams}\affiliation{University of Birmingham, Birmingham, United Kingdom}
\author{M.M.~Aggarwal}\affiliation{Panjab University, Chandigarh 160014, India}
\author{Z.~Ahammed}\affiliation{Variable Energy Cyclotron Centre, Kolkata 700064, India}
\author{J.~Amonett}\affiliation{Kent State University, Kent, Ohio 44242}
\author{B.D.~Anderson}\affiliation{Kent State University, Kent, Ohio 44242}
\author{M.~Anderson}\affiliation{University of California, Davis, California 95616}
\author{D.~Arkhipkin}\affiliation{Particle Physics Laboratory (JINR), Dubna, Russia}
\author{G.S.~Averichev}\affiliation{Laboratory for High Energy (JINR), Dubna, Russia}
\author{Y.~Bai}\affiliation{NIKHEF and Utrecht University, Amsterdam, The Netherlands}
\author{J.~Balewski}\affiliation{Indiana University, Bloomington, Indiana 47408}
% ...

\collaboration{STAR Collaboration}\noaffiliation


\begin{abstract}
The STAR Collaboration at RHIC reports measurements of azimuthal
correlations of high transverse momentum ($p_T$) charged hadrons in
Au+Au collisions at higher $p_T$ than reported previously.  As $p_T$
is increased, a narrow, back-to-back peak emerges above the decreasing
background, providing a clear dijet signal for all collision
centralities studied.  Using these correlations, we perform a systematic study
of dijet production and suppression in nuclear collisions, providing
new constraints on the mechanisms underlying partonic energy loss in
dense matter.
\end{abstract}


\maketitle

Nuclear collisions at high energy may produce conditions
sufficient for the formation of a deconfined plasma of quarks
and gluons \cite{qgpNPA}.  The high-density QCD matter
\cite{qgpNPA,WhitePapers} generated in these collisions can
be probed via propagation of hard scattered partons, which have
been predicted to lose energy in the medium primarily through gluon
bremsstrahlung \cite{gw1994,baierReview,gvwzReview,kovnerReview}.  The
medium alters the fragmentation of the parent partons, providing
experimental observables that are sensitive to the properties of QCD
matter at high density.

The study of high transverse momentum ($p_T$) hadron production in
heavy ion collisions at RHIC has yielded several novel results
\cite{jacobsWang}, including the strong suppression 
relative to p+p collisions of both inclusive
hadron yields \cite{phenixSupp, starSupp1, starSupp2, phobosSupp} and
back-to-back azimuthal ($\phi$) correlations \cite{starB2bPrl}.
Azimuthal correlations of high $p_T$ hadrons reflect the fragmentation of outgoing partons
produced dominantly in $2\!\rightarrow\!2$ hard scattering processes
(``dijets"
\cite{dijetRef}).  The back-to-back correlation strength has shown
sensitivity to the in-medium path length of the parton
\cite{star200v2Prl}, while the distribution of low $p_T$ hadrons
recoiling from a high $p_T$ particle is broadened azimuthally and
softened in central collisions \cite{starFuqiang}, qualitatively
consistent with dissipation of jet energy to the medium.  However,
those correlation measurements required large background
subtraction, and quantitative study of the properties of the away-side
jet has been limited. Previous correlation measurements also were
constrained to a \pT\ region in which the hadron flavor content and baryon fraction exhibit substantial differences from jet fragmentation in
elementary collisions \cite{phenixScaling,starK0Lambda,starHighpTPID}. 

In this Letter we present measurements of azimuthal correlations of
charged hadrons in Au+Au collisions at $\sqrt{s_{NN}}=200$ GeV over a much
broader transverse momentum range than previously reported. The $p_T$
range extends to the region where previous studies suggest that
particle production is dominated by jet fragmentation
\cite{phenixScaling,starK0Lambda,starHighpTPID}. Increasing $p_T$
reduces the combinatoric background and, for all centralities, reveals
narrow back-to-back peaks indicative of dijets. A quantitative study
of the centrality and $p_T$ dependence of dijet fragmentation may
provide new constraints on partonic energy loss and properties of the
dense medium (e.g. \cite{MullerRajagopal}).

The measurements were carried out with the STAR Experiment
\cite{starNim}, which is well-suited for azimuthal correlation studies
due to the full azimuthal ($2\pi$) coverage of its Time Projection
Chamber (TPC).  This analysis is based on 30M minimum-bias and
18M central Au+Au collisions at $\sqrt{s_{NN}}=200$ GeV, combining
the 2001 dataset with the high statistics dataset
collected during the 2004 run.  10M d+Au events collected in 2003 are
also included in the analysis.  Event and track selection are similar
to previous STAR high $p_T$ studies \cite{starSupp2,starDAu}.  This analysis used charged tracks from the primary vertex with pseudo-rapidity $|\eta|<1.0$.

As in our original studies of high $p_T$ azimuthal correlations
\cite{starB2bPrl}, transverse momentum-ordered jet-like correlations are measured by selecting
high $p_T$ {\it trigger} particles and studying the azimuthal
distribution of {\it associated} particles
$\bigl(\pTassoc\!<\!\pTtrig\bigr)$ relative to the trigger particle above a threshold $p_T$.
The trigger-associated technique facilitates jet studies in the
high-multiplicity environment of a heavy ion collision, where full jet
reconstruction using standard methods is difficult.  A particle may
contribute to more than one hadron pair in an event, both as trigger
and as associated particle, though for the high \pT\ ranges considered
here the rate of contribution to multiple pairs is small.  The pair yield is corrected for associated particle tracking efficiency, with uncertainty of 5\% that is highly correlated over the momentum range considered here. The effect of momentum resolution on the pair yield is estimated to be less than 1\% and no correction for it was applied.  A correction was also applied for non-uniform azimuthal acceptance, but not for the effects of
the single-track cut $|\eta|<1.0$. The single-track acceptance is independent of $p_T$ and uniform on $\eta$ for $p_T >3$ GeV/c and $|\eta|<1$. The near-side ($\dphi\sim0$)
correlated yield at large $|\deta|$ is negligible.

\begin{figure}[t]
\includegraphics*[width=3.3in]{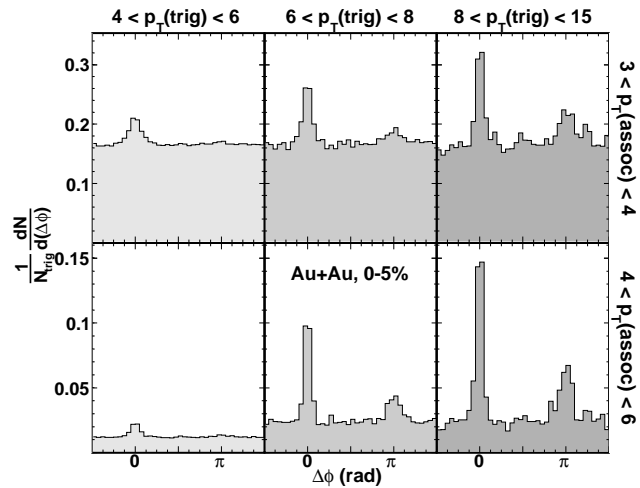} 
\caption{Azimuthal correlation histograms of high $p_T$ charged hadron pairs 
for 0-5\% Au+Au events, for various \pTtrig and \pTassoc ranges. 
In the lower left panel the yield is suppressed due to the constraint $\pTassoc<\pTtrig$.
All \pT\ values in this and succeeding figures have units GeV/c.}
\label{figure1}
\end{figure} 

Figure \ref{figure1} shows dihadron azimuthal distributions normalized
per trigger particle for central (0-5\%) Au+Au collisions.  \pTtrig\ increases
from left to right, and two \pTassoc\ ranges are shown.  The height of
the background away from the near- ($\dphi\sim0$) and away-side
($|\dphi|\sim\pi$) peaks, which is related to the inclusive yield, is
similar for different \pTtrig\ in each \pTassoc\ interval.  The
background level decreases rapidly as $\pTassoc$ is raised, {\it e.g.}
by an order of magnitude between the two rows in Fig.~1.  

Near-side peaks are seen in all panels and indicate larger yields for higher
\pTtrig\ at fixed \pTassoc. Such an increase in correlated yield is 
expected if the correlation is dominated by jet fragmentation, with
higher \pTtrig\ biasing towards higher $E_T$ jets. An away-side peak
is not apparent at the lowest \pTtrig, consistent with previous
studies of $\Delta\phi$ correlations in central Au+Au collisions in
similar \pTtrig\ and \pTassoc\ ranges \cite{starB2bPrl}. However, an away-side
peak emerges clearly above the background as \pTtrig\ is increased. The
narrow, back-to-back peaks are indicative of the azimuthally
back-to-back nature of dijets observed in elementary collisions.

\begin{figure}[t]
  {\includegraphics*[width=3.3in]{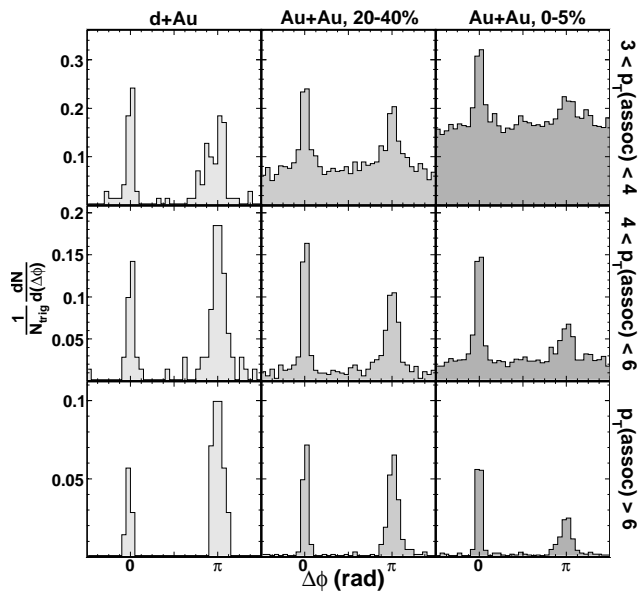}}
  \caption{Azimuthal correlation histograms of high $p_T$ charged hadrons 
for $8 < \pTtrig < 15$ GeV/c, for d+Au, 20-40\% Au+Au and 0-5\% Au+Au events. 
$\pTassoc$ increases from top to bottom.}\label{figure2}
\end{figure}

Figure \ref{figure2} shows the $\Delta\phi$ distributions for the
highest \pTtrig\ range in Fig.~1 ($8<\pTtrig <15$ GeV/c) for
mid-central (20-40\%) and central Au+Au collisions, as well as for d+Au collisions.
\pTassoc\ increases from top to bottom; for the highest \pTassoc\
(lower panels) the combinatorial background is negligible.  We observe
narrow correlation peaks in all
\pTassoc\ ranges. For each $\pTassoc$, the near-side peak shows similar correlation strength above background for
the three systems, while the away-side correlation strength decreases
from d+Au to central Au+Au. For $8<\pTtrig<15$ GeV/c and $\pTassoc\
>6$ GeV/c, a Gaussian fit to the away-side peak finds a width of
$\sigma_{\Delta\phi} =0.24\pm0.07$ for d+Au and $0.20\pm0.02$ and
$0.22\pm0.02$ for 20-40\% and 0-5\% Au+Au collisions, respectively. No
significant dependence of the widths on system or centrality is
observed.

\begin{figure}[t]
{\includegraphics*[width=3.3in]{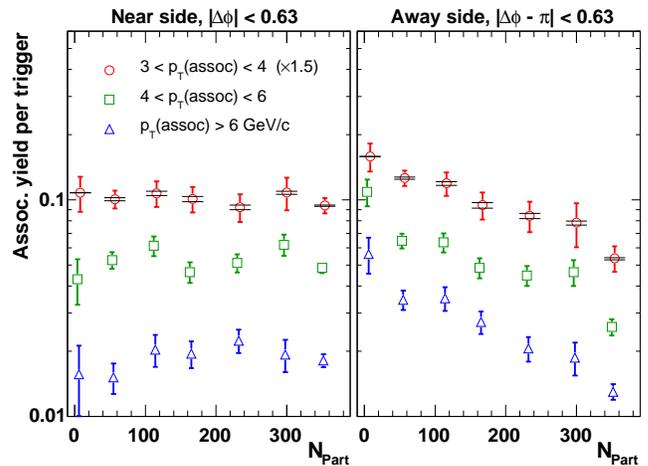}} 
\caption{(color online) Centrality dependence (number of participants $N_{\rm{Part}}$) of
near- and away-side yield in 200 GeV d+Au
(leftmost points) and Au+Au collisions, for $8<\pTtrig<15$ GeV/c and
various \pTassoc\ ranges. A semi-log scale is used and data for $3<\pTassoc<4$ GeV/c are scaled by 1.5 for clarity. The error bars are statistical. The horizontal bars for
$3<\pTassoc<4$ GeV/c show the systematic uncertainty due to background subtraction; it is negligible for higher \pTassoc.}
\label{figure3}
\end{figure}

To quantify the correlated near- and away-side yields we integrate
the area under the peaks (near-side $|\Delta\phi|<0.63$; away-side
$|\Delta\phi-\pi|<0.63$) and subtract the non-jetlike background.
In previous analyses at lower \pT, anisotropic (``elliptic") flow
contributed significantly to the measured two-particle correlation,
leading to large uncertainties in the extraction of jet-like yields
\cite{star200v2Prl,starFuqiang}. In this analysis, the background 
contribution due to elliptic flow is estimated using a function 
$B\,\cdot\,[1 + v_2\{\pTassoc\}\cdot v_2\{\pTtrig\}\cdot  \cos (2\Delta\phi)] $,
where the $v_2$ are extracted from standard elliptic flow analysis \cite{star200v2Prl} and $B$ is fitted to the region between the peaks ($0.63<|\Delta\phi|<2.51$),
 and is appreciable only for the lowest $\pTassoc$ range in
Fig.~2. The uncertainty in the magnitude of elliptic flow introduces a
small systematic uncertainty less than 5\% on the extracted associated yields (Figure \ref{figure3}).

Figure \ref{figure3} shows the centrality dependence of the near- and
away-side yield for the \pTtrig\ and \pTassoc\ ranges of
Figure~\ref{figure2}. The leftmost points in each panel correspond to
d+Au collisions, which we assume provide the reference distribution for jet fragmentation in vacuum. The near-side yields (left panel) show little
centrality dependence, while the away-side yields (right panel)
decrease with increasing centrality. The away-side centrality
dependence is similar to our previous studies of dihadron azimuthal
correlations for lower $p_T$ ranges \cite{starB2bPrl}.  Note that the yields in different \pTassoc\ bins for a given centrality may exhibit correlations due to their common trigger population.

\begin{figure}[t]
{\includegraphics*[width=3.3in]{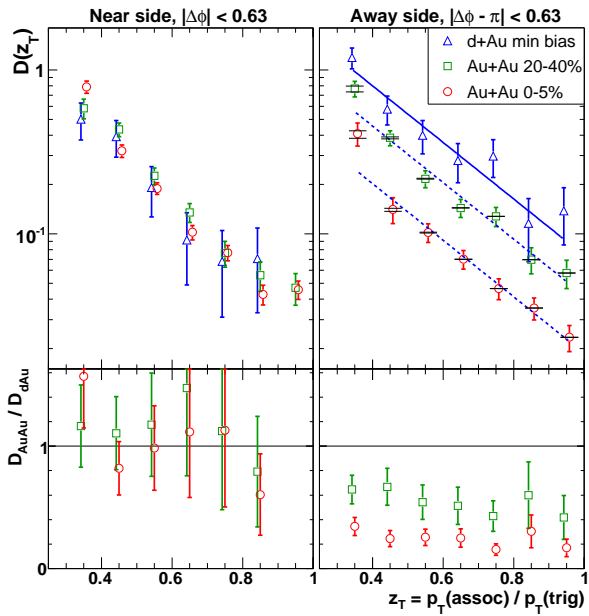}}
\caption{(color online) Upper panels: trigger-normalized charged hadron fragmentation
function $D(z_T)$ with $8 < \pTtrig < 15$ GeV/c, for near- (left) and
away-side (right) correlations in d+Au and Au+Au collisions at
$\sqrt{s_{NN}}$=200 GeV. Dashed and solid lines described in
text. Horizontal bars on away-side show systematic uncertainty due to
background subtraction. Lower panels: ratio of $D(z_T)$ for Au+Au
relative to d+Au. The error bars are statistical in all panels.}
\label{figure4}
\end{figure} 

The effect of the medium on dijet fragmentation can be explored in
more detail using the $p_T$ distributions of near- and away-side
associated hadrons. Fig.~4 shows the trigger-normalized fragmentation
function $D(z_T)$, where $z_T$ = $\pTassoc / \pTtrig$
\cite{xinNianFF}. $D(z_T)$ is measurable without direct knowledge
of the parton energy. The $z_T$ range shown in Fig.~4 corresponds to
the \pTassoc\ range for which dijets are observed above background
(see Fig.~2). The near-side distributions (left panels) are similar
over a broad range of $z_T$ for all three systems, consistent with
fragmentation in vacuum. 

The similarity of the near-side fragmentation patterns could arise
from small near-side energy loss due to a geometrical bias toward
shorter in-medium path lengths (``surface bias"), as generated in
several model calculations
\cite{dreesEtAl,DaineseLoizidesPaic,eskolaEtAl,Mueller02}. However, this
similarity could also result from energy-independent energy loss
generating a {\it partonic} energy distribution that is suppressed in Au+Au
but similar in shape to that in p+p collisions, with the lost
energy carried dominantly by low $p_T$ hadrons. A leading-twist calculation
of medium-modified dihadron fragmentation functions in similar
$\pTtrig$ and $\pTassoc$ intervals to those studied here
\cite{Majumder04} predicts a strong increase in near-side
associated yield for more central collisions, though no such increase is
observed in Figs.\@ 3 and 4.

The lower right panel of Figure~\ref{figure4} shows the ratio of
away-side $D(z_T)$ for 0-5\% and 20-40\% Au+Au relative to d+Au. The
ratio is approximately independent of $z_T$ for $z_T>0.4$, with yield
suppressed by a factor $0.25\pm0.06$ for 0-5\% Au+Au and $0.57\pm0.06$
for 20-40\% Au+Au collisions. The away-side suppression for central
collisions has similar magnitude to that for inclusive spectra
\cite{starSupp2}, though such similarity is not
expected {\it a priori} due to the different nature of the observable.
A model calculation based on BDMPS energy loss predicts a universal
ratio between away-side and inclusive suppression, with the away-side
yield more suppressed \cite{Mueller02}.

The solid line in Figure~\ref{figure4}, upper right panel, 
is an exponential function fit to the d+Au distribution, with the
dashed lines having the same exponential slope but magnitude scaled by
factors 0.57 and 0.25. This illustrates the similarity in
shape of $D(z_T)$ for different systems. As discussed for
Figure~\ref{figure2}, the width of the away-side azimuthal
distribution for high $p_T$ pairs is also independent of
centrality. To summarize our observations: strong away-side high $p_T$
hadron suppression is not accompanied by significant angular
broadening or modification of the momentum distribution for
$z_T>0.4$.

A calculation incorporating partonic energy loss through modification
of the fragmentation function \cite{xinNianFF} predicts the away-side
trigger-normalized fragmentation function to be suppressed uniformly
for $z_T>0.4$ in central Au+Au relative to p+p collisions, in
agreement with our measurement. However, the predicted magnitude of
the suppression is $\sim0.4$, weaker than the measured value
$0.25\pm0.06$.

Energy loss in matter could be accompanied by away-side azimuthal
broadening, due either to medium-induced acoplanarity of the parent
parton \cite{baierEtAl} or to dominance of the away-side yield
by medium-induced gluon radiation at large angle. An opacity expansion
calculation \cite{vitev05} predicts that the away-side yield for large
energy loss is dominated by fragments of the induced radiation, with a
strongly broadened azimuthal distribution up to $p_T\sim10$ GeV/c. No
azimuthal broadening of the away-side parent parton is predicted,
though its fragments are obscured by the greater hadron yield from
induced radiation. In contrast, we observe strong away-side
suppression without large azimuthal broadening. However, measurements
at $\pTassoc <1$ GeV/c do show an enhancement of the yield and
significant azimuthal broadening of the away-side peak
\cite{starFuqiang}.

Large energy loss is thought to bias the jet population generating the
high \pT\ {\it inclusive} hadron distribution towards jets produced near the
surface and directed outward
\cite{dreesEtAl,DaineseLoizidesPaic,eskolaEtAl,Mueller02}, which minimizes the
path length in the medium. For back-to-back dihadrons the total
in-medium path length is minimized by a different geometric bias,
towards jets produced near the surface but directed tangentially. A
model calculation \cite{daineseEtAlQM05} incorporating quenching
weights finds dihadron production dominated by such tangential pairs,
with yield suppression consistent with our measurements. Another
calculation based on quenching weights, which explicitly takes into
account the dynamical expansion of the medium \cite{Renk06}, also
reproduces the measured suppression but finds a significant
contribution from non-tangential jet pairs, due to the finite
probability to emit {\it zero} medium-induced gluons in finite path
length
\cite{xinNianFF,salgadoWiedemann} and to the rapid expansion and
dilution of the medium. In this model, the relative contribution
from the interior of the collision zone is larger for back-to-back
dihadrons than for inclusive hadron production.

In summary, we have measured new fragmentation properties of jets and
back-to-back dijets via high $p_T$ hadron correlations in
$\sqrt{s_{NN}}=200$ GeV d+Au and Au+Au collisions. We observe the
emergence at suppressed rate of a narrow back-to-back dijet peak in
central Au+Au collisions, which may enable the first {\it
differential} measurement of partonic energy loss. The
observation at high $p_T$ of strong suppression without modification
of the away-side azimuthal and $\pTassoc$ distributions is
in disagreement with several theoretical calculations. Other calculations reproduce aspects of these measurements but with somewhat different underlying mechanisms.  
New calculations are required to reconcile these differences and to
clarify the physics underlying our observations. We expect that
comparison of theory with the measurements reported here will provide
new insights into both the nature of partonic energy loss and the
properties of the medium generated in high energy nuclear collisions.

We thank the RHIC Operations Group and RCF at BNL, and the
NERSC Center at LBNL for their support. This work was supported
in part by the HENP Divisions of the Office of Science of the U.S.
DOE; the U.S. NSF; the BMBF of Germany; IN2P3, RA, RPL, and
EMN of France; EPSRC of the United Kingdom; FAPESP of Brazil;
the Russian Ministry of Science and Technology; the Ministry of
Education and the NNSFC of China; IRP and GA of the Czech Republic,
FOM of the Netherlands, DAE, DST, and CSIR of the Government
of India; Swiss NSF; the Polish State Committee for Scientific 
Research; STAA of Slovakia, and the Korea Sci. and Eng. Foundation.


\end{document}